\documentclass{IEEE_lsens}

\usepackage{textcomp}
\usepackage{graphicx}
\usepackage{subcaption}
\usepackage[export]{adjustbox}
\usepackage[noadjust]{cite}
\usepackage{xcolor}
\usepackage{array}
\usepackage[hidelinks]{hyperref}

\renewcommand{\thesubfigure}{(\alph{subfigure})}

\makeatletter
\def\ps@headings{%
  \let\@oddhead\@empty
  \let\@evenhead\@empty
  \let\@oddfoot\@empty
  \let\@evenfoot\@empty}

\def\ps@IEEEtitlepagestyle{%
  \def\@oddhead{%
    \hbox to\textwidth{%
      \raisebox{-9pt}[0pt][0pt]{%
        \parbox[t]{\textwidth}{%
          \raggedright\normalfont\sffamily
          \fontsize{8.0}{8.9}\selectfont
          \@IEEECOLORDEFtitle@blue{%
            \textbf{\textcopyright{} 2026 IEEE.} Personal use of this material is permitted. Permission from IEEE must be obtained for all other uses, in any current or future media, including reprinting/republishing this material for advertising or promotional purposes, creating new collective works, for resale or redistribution to servers or lists, or reuse of any copyrighted component of this work in other works.}%
        }%
      }\hss}%
  }%
  \let\@evenhead\@oddhead
  \let\@oddfoot\@empty
  \let\@evenfoot\@empty}

\def\@maketitle{\newpage
\bgroup\normalsize\par\vskip\IEEEtitletopspace\vskip\IEEEtitletopspaceextra\sffamily%
\rightskip\@flushglue\leftskip\z@skip\lineskiplimit 1pt\lineskip 1pt\relax
\vskip 1.95em%
\vskip 0.2em{\LARGE\bfseries\@IEEECOLORDEFtitle@blue{\@title}\par\mbox{}\\[-0.1\baselineskip]}\relax
\rightskip\@flushglue\leftskip\@flushglue
\centering
\@IEEEauthorabstractindextextbox{\ifCLASSOPTIONeditorial\else{\large\@author}\relax\fi%
\ifx\@IEEEspecialpapernotice\@empty\relax
\else
{\vskip 1.0em\par\@IEEEspecialpapernotice\par}\relax
\fi
\ifx\@IEEELSENSmanuscriptreceivedline\@empty\relax
\else
{\vskip 1.0em\par\normalfont\sffamily\scriptsize\@IEEELSENSmanuscriptreceivedline\par}\relax
\fi
\ifx\@IEEEtitleabstractindextext\@empty\relax
\else
{\vskip 0.8em\relax\par\@IEEEtitleabstractindextext\par}\relax
\fi}\relax
\par\egroup}
\makeatother

\begin{document}

\bstctlcite{BSTcontrol}

\markboth{}{}

\title{Toward Motion Capture Textiles: A Batteryless IMU Array Powered
and Networked Through a Conductive-Fabric Bus}

\author{\IEEEauthorblockN{Ryusei~Miyakawa\IEEEauthorrefmark{1}\IEEEauthorieeemembermark{1}
and~Akihito~Noda\IEEEauthorrefmark{1}\IEEEauthorieeemembermark{2}}%
\IEEEauthorblockA{\IEEEauthorrefmark{1}School of Systems Engineering,
Kochi University of Technology, Kochi, 782-8502, Japan\\
\IEEEauthorieeemembermark{1}Student Member, IEEE;
\IEEEauthorieeemembermark{2}Member, IEEE}%
\thanks{Corresponding author: A. Noda (e-mail: noda.akihito@kochi-tech.ac.jp).}%
}

\IEEELSENSmanuscriptreceived{Accepted manuscript. Accepted for publication in IEEE Sensors Letters.}

\IEEEtitleabstractindextext{%
\begin{abstract}[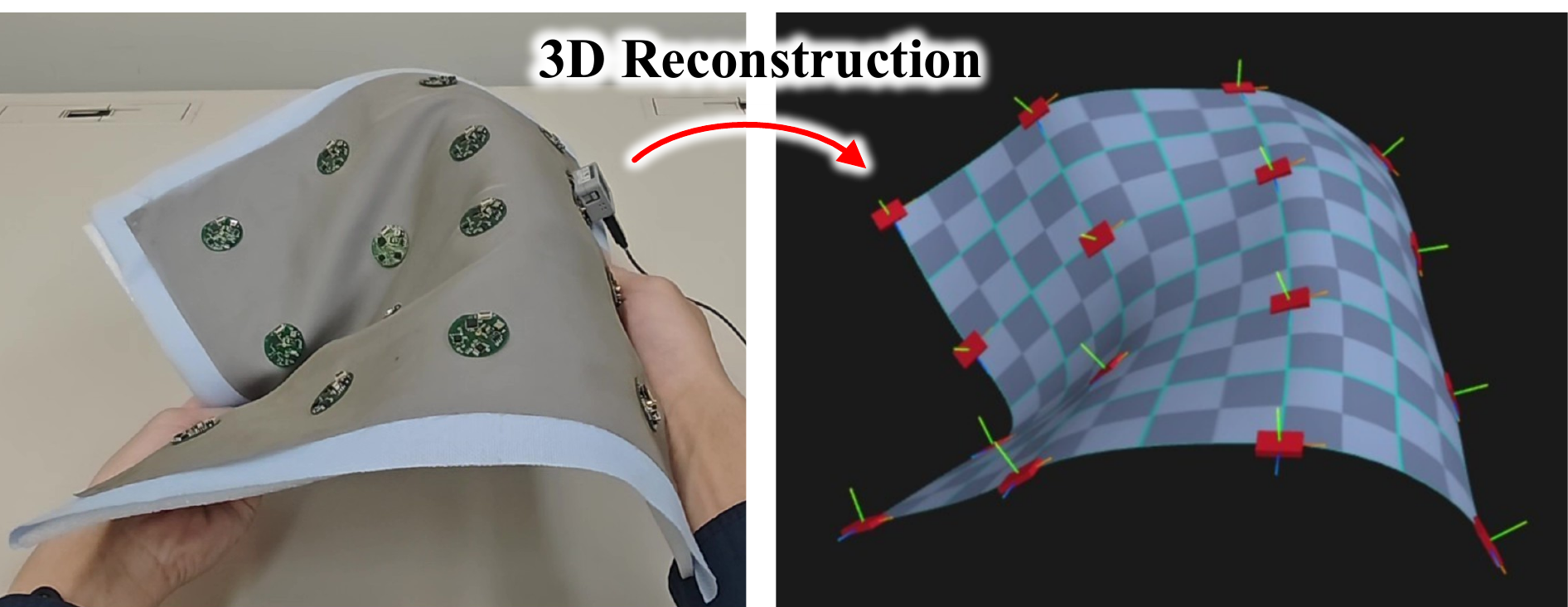]
Motion capture is increasingly important in healthcare, rehabilitation, and sports applications.
However, conventional vision-based systems require camera infrastructure and line-of-sight conditions, limiting their use in daily environments.
This paper proposes a location-independent platform for motion capture, based on an array of IMU (inertial measurement unit) sensor nodes without individual wiring or onboard batteries, enabled by two-dimensional communication and power delivery over conductive fabric.
Pipeline processing in each sensor node, combined with a token-passing protocol, enables collision-free and stable data acquisition at 50 Hz.
Real-time three-dimensional reconstruction of flexible objects is further demonstrated by applying Hermite spline interpolation to orientation data from grid-aligned IMUs.
Against a cylinder of known radius, the reconstructed surface deviates by 2.60 mm RMS over the 225 mm span of the array.
The reconstruction is formulated for the quasi-static regime, and this letter empirically validates it in that regime.
The hardware platform and communication protocol implemented in this work enable further investigation toward dynamic motion capture textiles.
\end{abstract}

}

\maketitle

\section{Introduction}
Motion capture systems are increasingly used for quantitative evaluation of human movement in healthcare~\cite{rihabiri} and sports science~\cite{sport}.
Continuous measurement of body posture and motion provides kinematic information such as gait parameters and subtle changes in movement~\cite{hokou}, and discriminating walking styles at the level of individual joints requires inertial data from many points on the body~\cite{semwal}.
Such information can support applications including early detection of motor disorders~\cite{pa-kin} and fall-risk assessment~\cite{tentou}, for which inertial sensing combined with learned models has been reported~\cite{gaud}.

Existing systems are broadly classified as external-sensor-based or wearable.
Optical motion capture achieves high accuracy but requires dedicated camera infrastructure and controlled environments~\cite{kougaku},~\cite{kougaku_2}, and it is susceptible to occlusion.
Hybrid optical--inertial systems mitigate occlusion~\cite{IMU_kougaku} at the cost of system complexity and user burden.

Wearable systems using only IMUs remove the dependence on external cameras~\cite{IMU_1,IMU_3,IMU_4}, supported by mature orientation filters~\cite{madgwick},~\cite{mahony} and by products such as Sony's mocopi~\cite{mocopi}.
They still require either wireless nodes with individual batteries~\cite{musen} or wired connections between nodes~\cite{wire}.
Batteries add weight and maintenance as the node count grows, and cabling across the body complicates setup and daily use.

Conductive textiles can instead carry power and data over a shared two-dimensional medium~\cite{i2we}, and we propose on that basis a shape-sensing textile with inertial measurement unit (IMU) sensor nodes placed on conductive fabric.
The fabric, woven with copper--nickel (Cu--Ni)-plated threads, is a two-dimensional shared medium for power and data, eliminating inter-node wiring and onboard batteries while allowing sensors anywhere on the surface.

Our previous work~\cite{2D-communication} established the signaling scheme used here.
It models the textile line as a second-order circuit, derives the critical-damping condition for comparator-based recovery, and demonstrates one-way broadcast from a single transmitter to 28 batteryless receivers, none of which transmits.
The present letter contributes (i) the integration of that scheme with an IMU and a microcontroller on a single 0.9 g board powered from the fabric, (ii) a token-passing protocol with per-node pipelining of sensing, orientation computation and transmission, sustaining a 50 Hz frame rate across 16 such nodes, and (iii) real-time surface reconstruction from the array, evaluated against known geometry.
As a step toward wearable motion capture, this letter validates the platform at the textile level; no on-body measurement is claimed.

\section{Hardware and Communication Implementation}
\label{section_2}

\subsection{Conductive-Fabric Platform}
The proposed system uses conductive fabric as a two-dimensional shared medium for power delivery and data communication.
As shown in Fig.~\ref{fig:sub_a}, IMU sensor nodes are placed directly on the fabric, and a master node supplies power to the fabric bus while collecting posture data from the sensor nodes.
The collected data are then transferred to a PC for real-time shape reconstruction.

\begin{figure}[!t]
    \centering
    {\setlength{\fboxsep}{0pt}\setlength{\fboxrule}{0.4pt}%
    \stepcounter{figure}\setcounter{subfigure}{0}%
    \begin{minipage}[c]{0.045\linewidth}
        \refstepcounter{subfigure}\label{fig:sub_a}\thesubfigure
    \end{minipage}%
    \begin{minipage}[c]{0.94\linewidth}
        \includegraphics[width=\linewidth]{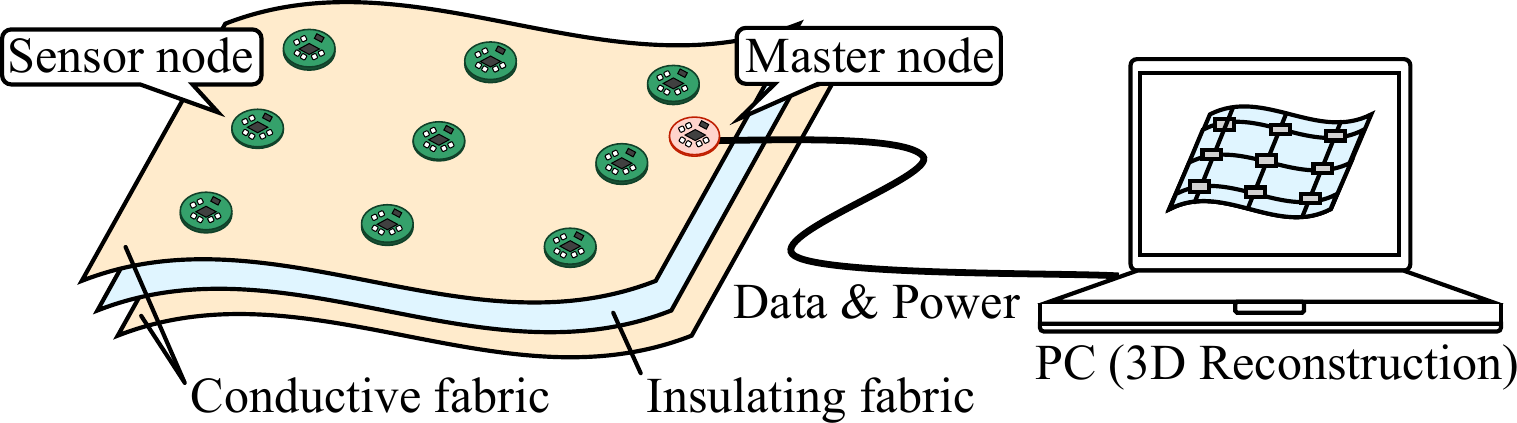}
    \end{minipage}\\[2pt]
    \begin{minipage}[c]{0.045\linewidth}
        \refstepcounter{subfigure}\label{fig:sub_b}\thesubfigure
    \end{minipage}%
    \begin{minipage}[c]{0.94\linewidth}
        \includegraphics[width=\linewidth]{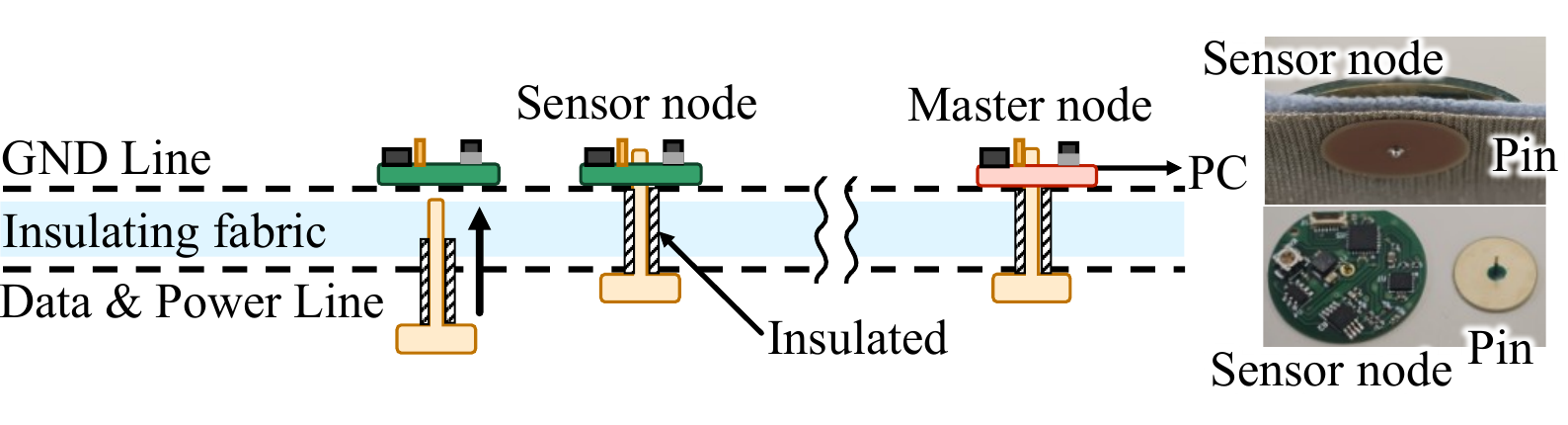}
    \end{minipage}}
    \addtocounter{figure}{-1}%
    \caption{Batteryless IMU sensor array on a conductive-fabric bus.
    (a) System overview. (b) Cross-section of the sensor and master nodes, with photographs of the custom node and contact pin.
    The master supplies power to the bottom conductive layer, through which the surface-mounted nodes communicate via conductive pins.}
    \label{fig:main}
\end{figure}

\begin{table}[!t]
\caption{System Specifications}
\label{tab:spec}
\centering
\footnotesize
\renewcommand{\arraystretch}{1.08}
\begin{tabular}{@{}>{\raggedright}p{0.27\linewidth}p{0.65\linewidth}@{}}
\hline
\multicolumn{2}{@{}l}{\textit{Conductive-fabric bus}}\\
\hline
Conductive textile & DW-372N\textsuperscript{a}, Cu--Ni plated fiber, 0.15 mm\\
Sheet resistance & 0.3 $\Omega$/sq (four-point method)\\
Insulating textile & Cotton 55\%/polyester 45\%, 0.3 mm\\
Bus / array geometry & $300\times300$ mm; $4\times4$ grid, pitch $L=75$ mm\\
Bus rate / frame rate & 1 Mbps (UART on DC) / 50 Hz for 16 nodes\\
\hline
\multicolumn{2}{@{}l}{\textit{Sensor node} ($\times$16)}\\
\hline
PCB & 22 mm diameter, 0.9 g, batteryless\\
IMU & ICM-42670-P\textsuperscript{b}, 6-axis\\
MCU & ATmega328P\textsuperscript{c}, 8 MHz\\
Orientation filter & Madgwick, gain 0.12 rad/s, 50 Hz\\
Supply / current & 3.3 V / 10.7 mA per node\\
AC decoupling & 1 mH (CB2518T102K\textsuperscript{d}), DCR 24 $\Omega$, $I_\mathrm{sat}$ 25 mA\\
Comparator hyst. & 110 mV\\
\hline
\multicolumn{2}{@{}l}{\textit{Master node} ($\times$1)}\\
\hline
PCB / weight & $24\times28$ mm, 9.1 g with the MCU module\\
MCU module & M5Atom Matrix\textsuperscript{e} (ESP32-PICO\textsuperscript{f})\\
Power & USB Type-C bus-powered\\
AC decoupling & 8.3 \textmu H (NFZ32BW520HN\textsuperscript{g}), DCR 0.26 $\Omega$\\
\hline
\multicolumn{2}{@{}p{0.93\linewidth}@{}}{\vspace{1pt}\scriptsize
DCR: DC resistance. \textsuperscript{a}MAC Corp., Osaka, Japan. \textsuperscript{b}TDK InvenSense, San Jose, CA, USA. \textsuperscript{c}Microchip Technology, Chandler, AZ, USA. \textsuperscript{d}Taiyo Yuden, Tokyo, Japan. \textsuperscript{e}M5Stack, Shenzhen, China. \textsuperscript{f}Espressif Systems, Shanghai, China. \textsuperscript{g}Murata Mfg., Kyoto, Japan.}
\end{tabular}
\end{table}

The battery-free design keeps each node below 1 g, which matters because sensor mass mechanically loads the measured object.
Nodes connect at arbitrary positions, so the sensor density can be matched to the target.

In the three-layer stack of Fig.~\ref{fig:sub_b} the top Cu--Ni-plated layer is a common ground and the bottom one a shared signal and power layer, so all nodes are connected in parallel regardless of mounting position.

Each node attaches through a pin-and-socket interface (Fig.~\ref{fig:sub_b}).
Ground (GND) pads contact the top layer, while insulated pins penetrate the stack to the bottom signal/power layer.

\subsection{Sensor Node Design}
Each node carries a 6-axis IMU, a microcontroller unit (MCU) and a transceiver circuit on a 22 mm printed circuit board (PCB), and is powered through the fabric rather than by an onboard battery (Table~\ref{tab:spec}).

Power delivery and bidirectional communication share the signal/power layer.
DC power from the master passes through an inductor that decouples the data path.
Universal asynchronous receiver/transmitter (UART) data are coupled to the same layer as transient pulses generated at the UART edges, and are recovered at the receiving node by a hysteresis comparator.

The decoupling elements of all connected nodes appear in parallel, and their combined inductance $L_\parallel$ must satisfy the critical-damping condition of~\cite{2D-communication}.
The master carries the whole array current, so its element is set close to the target $L_\parallel$ to keep its DC resistance (DCR) low.
A sensor node passes only its own 10.7 mA and can afford 1 mH, whose 24 $\Omega$ DCR costs 257 mV.
Sixteen such nodes in parallel still present 62.5 \textmu H against the master's 8.3 \textmu H, so $L_\parallel$ stays near its design value as nodes are added.

All parameters required to reproduce the system are given in Table~\ref{tab:spec}.

\subsection{Data Acquisition Protocol}

Each node fuses the three-axis acceleration and three-axis angular velocity of its IMU into a quaternion using a Madgwick complementary filter on the MCU (Table~\ref{tab:spec}), and transmits only the resulting quaternion.
No smoothing or outlier rejection is applied to it, so the chain from raw inertial samples to the reconstructed surface contains no tuned preprocessing stage.
Because six-axis IMUs are used without a magnetometer, no absolute heading reference is available; the consequence is discussed in Section~\ref{sec:limitations}.
Token access reads the nodes sequentially, so samples within a frame are skewed by up to one frame period, 20 ms at 50 Hz, which displaces a node by 20 \textmu m per mm/s of deformation speed.
This term is absent from the static evaluation of Section~\ref{sec:eval} and is one reason the quantitative accuracy is established there.

Because all nodes are connected in parallel, simultaneous transmissions would collide, so a token-passing protocol assigns transmission rights sequentially.

\begin{figure}
    \centering
    \includegraphics[width=0.85\linewidth]{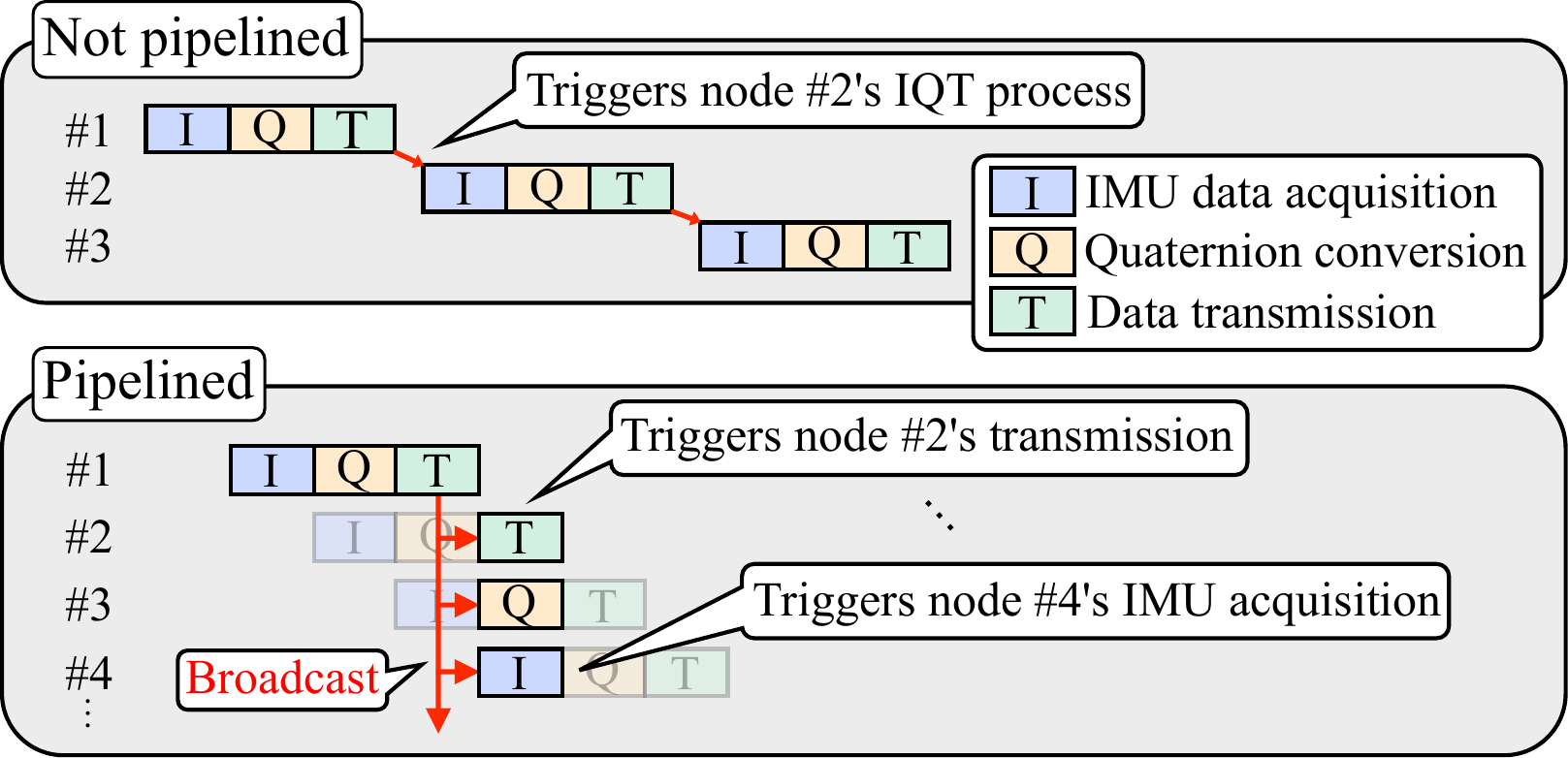}
    \caption{Communication sequence with and without per-node pipelining, showing the data transmission (T), IMU acquisition (I) and quaternion conversion (Q) of each node. The interval saved per node is what permits a 50 Hz frame rate with 16 nodes.}
    \label{sequence}
\end{figure}

As shown in Fig.~\ref{sequence}, each node ends its slot by broadcasting a token naming its successor, which acknowledges and transmits in turn.
The master monitors the bus and re-issues an unacknowledged token to the following node.
Arbitration is thus distributed for speed, while recovery remains centralized.

Sensing, computation and communication are pipelined within each node.
Because every transmission is broadcast, a node begins its IMU acquisition and quaternion conversion two slots before its turn, so current data are ready when the token arrives.

The number of nodes is bounded both by time and by supply voltage.
Each slot carries 12 bytes: eight for the quaternion, at two bytes per component, one for a cyclic redundancy check (CRC), and three of protocol overhead.
Including the UART start and stop bits a byte occupies 10 \textmu s at 1 Mbps, so only 120 \textmu s of the measured 1269 \textmu s interval is bus time, and the frame rate is governed by processing within each node rather than by the bus.
At 50 Hz the array is limited to about 16 nodes.
A lower frame rate admits proportionally more nodes, and faster firmware would raise the limit without any change to the bus.
The second bound keeps the farthest node above its minimum operating voltage after the drops along the bus and across its own decoupling element, giving about 24 nodes here.
This follows from powering the nodes directly at 3.3 V rather than from the architecture itself.
Driving the bus higher and regulating locally would remove it, at the cost of the size and mass a regulator adds to every node.
Frame rate, node count and node mass thus form a trade-off, and which bound is reached first depends on the operating point.

\section{Cloth Shape Reconstruction}
\label{section_3}

\subsection{Reconstruction Principle}
Conventional IMU-based motion capture maps orientation data onto a predefined kinematic skeleton, which a fabric does not possess.
Its shape is instead reconstructed by combining IMU orientations with the known grid arrangement of the nodes.
Double integration of acceleration is unsuitable, as drift accumulates and distorts the relative positions.
Rather than solve a full cloth-mechanics problem~\cite{chebyshev},~\cite{mass-spring}, we assume negligible in-plane stretching and a constant internode distance, which is light enough for real-time visualization at 50 Hz.

\subsection{Node Position and Surface Estimation}
Adjacent node positions are estimated from fixed-length displacement vectors whose directions follow an intermediate orientation $q_{i,j}$, computed by spherical linear interpolation (Slerp)~\cite{kyuumen} so that neither endpoint alone determines the local tangent.
The position of node $j$ is estimated from that of node $i$ as
\begin{equation}
    \mathbf{p}_j
    =
    \mathbf{p}_i
    +
    q_{i,j}
    \left(
    L \mathbf{v}_{i,j}
    \right)
    q_{i,j}^{-1},
    \label{eq:relative_position}
\end{equation}
where $L$ is the internode distance and $\mathbf{v}_{i,j}$ the unit direction from node $i$ to node $j$ in the local grid frame.
Positions propagate outward from the grid center, averaging estimates from already-computed neighbors, and a continuous surface follows by cubic Hermite interpolation of the coordinates and the tangents derived from the node orientations.

\section{Experiments and Evaluation}
\label{sec:eval}

\begin{figure}[!tb]
\centering
\includegraphics[width=\linewidth]{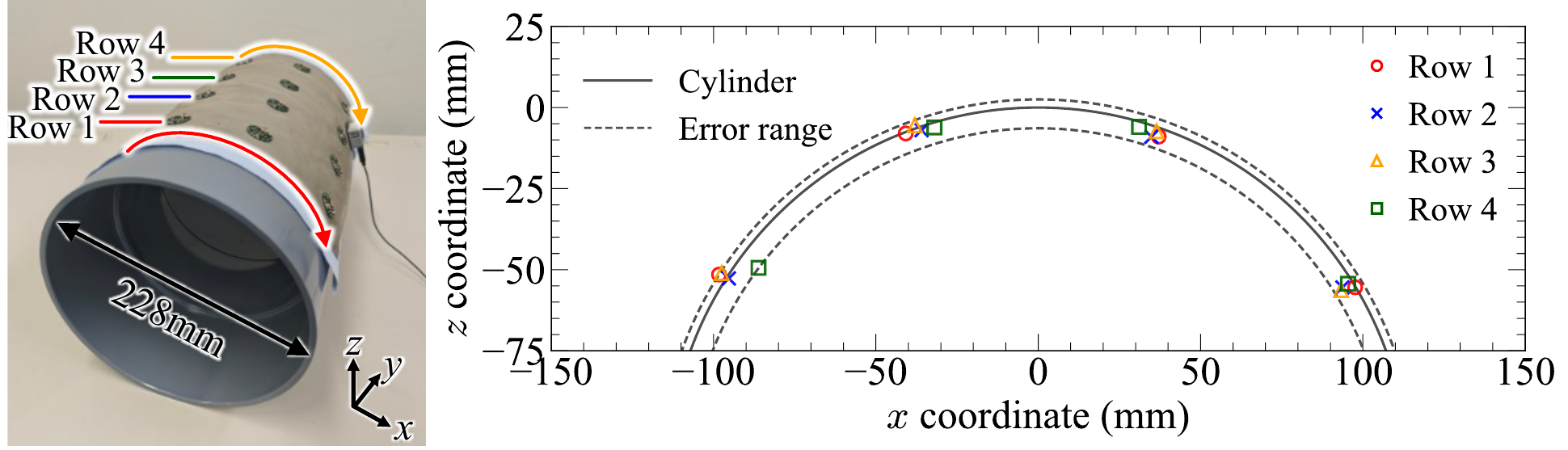}
\caption{Static accuracy against a 228 mm diameter cylinder. Estimated coordinates of all 16 nodes, grouped by row and projected onto the plane normal to the cylinder axis, overlaid on the true arc. Dashed curves bound the deviation.}
\label{fig:cylerr}
\end{figure}

\begin{figure}
    \centering
    \includegraphics[width=0.95\linewidth]{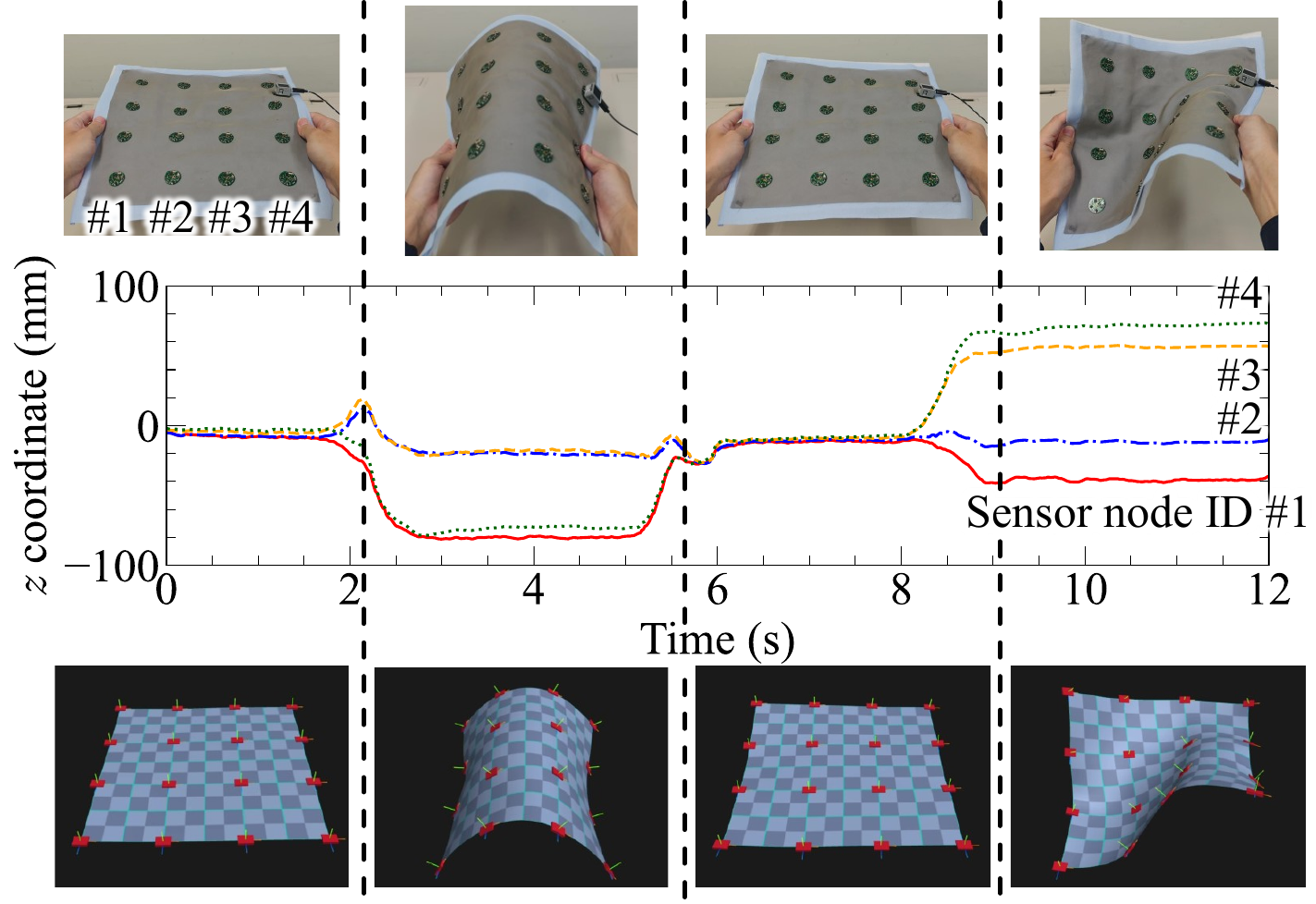}
    \caption{Reconstruction using 16 IMUs in a 4$\times$4 grid. Photographs, reconstructed surfaces, and the $z$ coordinates of the four nodes along one edge (IDs 1--4) about the fabric center. The fabric is flat ($0-2$, $6-8\,\mathrm{s}$), arched ($2-6\,\mathrm{s}$), and complexly deformed ($8-12\,\mathrm{s}$).}
    \label{fig:IMU_4x4}
\end{figure}

\subsection{Experimental Setup}
The prototype of Table~\ref{tab:spec} sustains a 50 Hz frame rate, above the bandwidth of the deformations considered here~\cite{50Hz}, with the master connected to a PC for reconstruction and logging.

The six-axis IMUs have no magnetometer and so provide no absolute yaw reference.
Before each measurement the fabric was therefore laid flat on a horizontal table, where all 16 nodes face the same yaw direction, and each IMU orientation at that instant was taken as the reference for the reconstruction.
Ground-truth geometry was obtained by wrapping the fabric around a rigid cylinder of 228 mm diameter, so that the true node coordinates follow analytically from the cylinder geometry.
No human participants were involved, and the results below come from a single recording.

\subsection{Communication Performance}
The 16 nodes draw 171 mA in total, leaving the farthest node at 3.06 V on the bus and 2.80 V after its own decoupling element, both above the 2.7 V minimum.
There the received pulses had an amplitude of about 400 mV against a comparator hysteresis of about 110 mV, a margin of about 11 dB.
Over 10 min of continuous operation the master completed 29\,692 ID1-to-ID1 cycles with one corrupted frame, an error rate of $3.4\times10^{-5}$.
The mean period was 20.206 ms with a standard deviation of 0.127 ms, so the schedule is deterministic to 0.6\%.
The sheet resistance of the fabric therefore affects the supply margin more than it affects signaling.

\subsection{Reconstruction Accuracy}
The 16 estimated node positions deviate from the true arc by between $-6.39$ mm and $+2.55$ mm, a positive deviation lying outside the cylinder (Fig.~\ref{fig:cylerr}).
Over the $31\times31$ grid points used internally, the surface deviates with a mean of $-1.84$ mm and a standard deviation of 1.84 mm, an RMS error of 2.60 mm or 1.2\% of the 225 mm span of the array.
The reconstruction thus sits slightly inside the true surface, and the spread grows with distance from the grid center.
The innermost four nodes deviate by $-3.37$ mm to $+0.96$ mm, while the outermost four span the full $-6.39$ mm to $+2.55$ mm and set the envelope of the array.
Using only the four corner nodes raises the surface standard deviation from 1.84 mm to 4.70 mm, quantifying the benefit of the denser grid.

\subsection{Quasi-Static Deformation Tracking}
Rather than a further quantitative assessment, which Section~\ref{sec:eval}-C provides statically, this subsection shows how the reconstruction behaves while the fabric is deformed. A dynamic assessment would require optical motion capture as ground truth and is left as future work.
Fig.~\ref{fig:IMU_4x4} plots the $z$ coordinates of the four nodes along one edge (IDs 1--4) about the fabric center.
In the flat states the nodes stay within about 2 mm of $z=0$.
Between 2 and 6 s the fabric is arched, the inner nodes (IDs 2 and 3) staying near the origin while the outer ones reach about $-80$ mm.
It then returns to flat without recalibration, and from 8 to 12 s each node traces an independent trajectory under complex deformation, throughout without frame loss.
Excursions of roughly 80 mm develop over about 1 s, so node speeds stay below 0.1 m/s and accelerations are of order $0.1g$.
The recording therefore lies in the assumed quasi-static regime.

\subsection{Limitations}
\label{sec:limitations}
Four limitations should be noted.
Each node derives orientation from a filter that uses gravity as attitude reference, so deformation rapid enough for linear acceleration to rival gravity lies outside the validated quasi-static range.
Because the nodes have no absolute heading reference, yaw drift accumulates, and the flat-state error grew from 1.4 mm to 29.5 mm over 1 min.
Long-term use would therefore require periodic recalibration, or a magnetometer to supply the missing reference.
The surface is assumed inextensible.
Since the cylinder used above is developable, accuracy should degrade on non-developable surfaces and under in-plane strain.
Finally, validation is at the textile level only.

\section{Conclusion}
This paper presented a batteryless and wire-free shape reconstruction system using conductive fabric.
Using the fabric as a shared power and data bus, and superimposing data on DC power via bias-tee circuits, eliminated the need for individual batteries and wiring at each sensor node.
A token-passing protocol combined with pipelined data processing achieved collision-free real-time data acquisition at 50 Hz in a 4$\times$4 sensor network.
Against a cylinder of known radius the reconstructed surface deviated by 2.60 mm RMS over the 225 mm span of the array, obtained without a predefined skeletal model.

The 50 Hz figure is a property of the acquisition system rather than a validated dynamic accuracy.
Motion capture in the full sense will require a reconstruction formulated for dynamic deformation and its characterization against optical motion capture.
Because the bus delivers power and data to any node type at any position, heterogeneous nodes carrying force or pressure sensors can also be added toward zero-moment-point (ZMP) and ground-reaction-force estimation~\cite{maske}.
These, with on-body measurement, are the intended next steps.

\section*{Acknowledgment}
\addcontentsline{toc}{section}{Acknowledgment}
\scriptsize
This research was supported in part by JSPS KAKENHI JP25K03109 and JST CRONOS, Japan, Grant Number JPMJCS25N4.
The authors used Claude (Anthropic) for English language editing and Gemini (Google) for programming assistance in developing the visualization tool shown in Fig.~\ref{fig:IMU_4x4}.
\normalsize

\bibliographystyle{IEEEtran}
\bibliography{reference_corrected}

\begin{thebibliography}{10}
\providecommand{\url}[1]{#1}
\csname url@samestyle\endcsname
\providecommand{\newblock}{\relax}
\providecommand{\bibinfo}[2]{#2}
\providecommand{\BIBentrySTDinterwordspacing}{\spaceskip=0pt\relax}
\providecommand{\BIBentryALTinterwordstretchfactor}{4}
\providecommand{\BIBentryALTinterwordspacing}{\spaceskip=\fontdimen2\font plus
\BIBentryALTinterwordstretchfactor\fontdimen3\font minus
  \fontdimen4\font\relax}
\providecommand{\BIBforeignlanguage}[2]{{%
\expandafter\ifx\csname l@#1\endcsname\relax
\typeout{** WARNING: IEEEtran.bst: No hyphenation pattern has been}%
\typeout{** loaded for the language `#1'. Using the pattern for}%
\typeout{** the default language instead.}%
\else
\language=\csname l@#1\endcsname
\fi
#2}}
\providecommand{\BIBdecl}{\relax}
\BIBdecl

\bibitem{rihabiri}
L.~Carcreff \emph{et~al.}, ``Validity, reliability and interpretability of an
  {IMU}-based system to measure {3D} lower limb kinematics of patients with
  heterogeneous gait disorders,'' \emph{Sensors}, vol.~26, 2026, art. no. 1746.

\bibitem{sport}
E.~van~der Kruk and M.~M. Reijne, ``Accuracy of human motion capture systems
  for sport applications; state-of-the-art review,'' \emph{European Journal of
  Sport Science}, vol.~18, no.~6, pp. 806--819, 2018.

\bibitem{hokou}
J.~C.~M. Schlachetzki \emph{et~al.}, ``Wearable sensors objectively measure
  gait parameters in {Parkinson's} disease,'' \emph{PLOS ONE}, vol.~12, no.~10,
  2017, art. no. e0183989.

\bibitem{semwal}
V.~B. Semwal \emph{et~al.}, ``Pattern identification of different human joints
  for different human walking styles using inertial measurement unit ({IMU})
  sensor,'' \emph{Artificial Intelligence Review}, vol.~55, pp. 1149--1169,
  2022.

\bibitem{pa-kin}
M.~Gregor{\v{c}}i{\v{c}} and D.~Georgiev, ``The usefulness of wearable sensors
  for detecting freezing of gait in {Parkinson's} disease: A systematic
  review,'' \emph{Sensors}, vol.~25, no.~16, 2025, art. no. 5101.

\bibitem{tentou}
Z.~K. Lim \emph{et~al.}, ``Fall risk prediction using temporal gait features
  and machine learning approaches,'' \emph{Frontiers in Artificial
  Intelligence}, vol.~7, 2024, art. no. 1425713.

\bibitem{gaud}
N.~Gaud \emph{et~al.}, ``{FIBiLS}: Fall detection of healthy elderly using
  {IMU} sensor and {BiLSTM} model,'' \emph{IEEE Sensors Journal}, vol.~25,
  no.~19, pp. 37\,124--37\,131, 2025.

\bibitem{kougaku}
B.~Adlou, C.~Wilburn, and W.~Weimar, ``Motion capture technologies for athletic
  performance enhancement and injury risk assessment: A review for multi-sport
  organizations,'' \emph{Sensors}, vol.~25, 2025, art. no. 4384.

\bibitem{kougaku_2}
H.~Joo, T.~Simon, and Y.~Sheikh, ``Total capture: A {3D} deformation model for
  tracking faces, hands, and bodies,'' in \emph{Proceedings of the IEEE
  Conference on Computer Vision and Pattern Recognition}, 2018, pp. 8320--8329.

\bibitem{IMU_kougaku}
P.~Furgale, J.~Rehder, and R.~Siegwart, ``Unified temporal and spatial
  calibration for multi-sensor systems,'' in \emph{2013 IEEE/RSJ International
  Conference on Intelligent Robots and Systems}, 2013, pp. 1280--1286.

\bibitem{IMU_1}
H.~Yang \emph{et~al.}, ``Reliability enhancement method of attitude estimation
  for wearable motion-capture systems in human lower limb rehabilitation,''
  \emph{IEEE Sensors Journal}, vol.~23, no.~21, pp. 26\,677--26\,690, 2023.

\bibitem{IMU_3}
T.~von Marcard \emph{et~al.}, ``Sparse inertial poser: Automatic {3D} human
  pose estimation from sparse {IMUs},'' \emph{Computer Graphics Forum},
  vol.~36, no.~2, pp. 349--360, 2017.

\bibitem{IMU_4}
Y.~Huang \emph{et~al.}, ``Deep inertial poser: Learning to reconstruct human
  pose from sparse inertial measurements in real time,'' \emph{ACM Transactions
  on Graphics}, vol.~37, no.~6, pp. 1--15, 2018.

\bibitem{madgwick}
S.~O.~H. Madgwick, ``An efficient orientation filter for inertial and
  inertial/magnetic sensor arrays,'' University of Bristol, Tech. Rep., 2010.

\bibitem{mahony}
R.~Mahony, T.~Hamel, and J.-M. Pflimlin, ``Nonlinear complementary filters on
  the special orthogonal group,'' \emph{IEEE Transactions on Automatic
  Control}, vol.~53, no.~5, pp. 1203--1218, 2008.

\bibitem{mocopi}
\BIBentryALTinterwordspacing
{Sony Corporation}, ``{mocopi} developer site,'' 2023, accessed: 2026-05-04.
  [Online]. Available: \url{https://www.sony.co.jp/en/Products/mocopi-dev/en/}
\BIBentrySTDinterwordspacing

\bibitem{musen}
F.~Gr{\"u}tzmacher \emph{et~al.}, ``{fastSW}: Efficient piecewise linear
  approximation of quaternion-based orientation sensor signals for motion
  capturing with wearable {IMUs},'' \emph{Sensors}, vol.~21, no.~15, 2021, art.
  no. 5180.

\bibitem{wire}
A.~Tsukamoto, K.~Mase, and Y.~Enokibori, ``Integrated utilization of
  {IMU}-based human activity recognition datasets across varied configurations
  using {MIG HAR} dataset,'' \emph{International Journal of Activity and
  Behavior Computing}, vol. 2024, no.~2, pp. 1--21, 2024.

\bibitem{i2we}
A.~Noda and H.~Shinoda, ``Inter-{IC} for wearables ({I2We}): Power and data
  transfer over double-sided conductive textile,'' \emph{IEEE Transactions on
  Biomedical Circuits and Systems}, vol.~13, no.~1, pp. 80--90, 2019.

\bibitem{2D-communication}
A.~Noda, ``{UART} for wearables ({U4We}): {DC} power and carrierless signal
  transfer over conductive textiles,'' \emph{IEEE Transactions on Circuits and
  Systems---II: Express Briefs}, 2026, doi:
  \href{https://doi.org/10.1109/TCSII.2026.3735640}
  {10.1109/TCSII.2026.3735640}.

\bibitem{chebyshev}
M.~Magri and P.~Ciarletta, ``Mathematics meets the fashion industry on path to
  product innovation and sustainability,'' \emph{Proceedings of the Royal
  Society A: Mathematical, Physical and Engineering Sciences}, vol. 479, 2023,
  art. no. 20220715.

\bibitem{mass-spring}
X.~Provot, ``Deformation constraints in a mass-spring model to describe rigid
  cloth behaviour,'' in \emph{Proceedings of Graphics Interface '95}, 1995, pp.
  147--154.

\bibitem{kyuumen}
K.~Shoemake, ``Animating rotation with quaternion curves,'' in
  \emph{Proceedings of the 12th Annual Conference on Computer Graphics and
  Interactive Techniques}, ser. {SIGGRAPH} '85, vol.~19, no.~3, 1985, pp.
  245--254.

\bibitem{50Hz}
J.~Gonz{\'a}lez-Alonso \emph{et~al.}, ``Custom {IMU}-based wearable system for
  robust 2.4 {GHz} wireless human body parts orientation tracking and {3D}
  movement visualization on an avatar,'' \emph{Sensors}, vol.~21, 2021, art.
  no. 6642.

\bibitem{maske}
N.~Maske \emph{et~al.}, ``{ZMP} estimation from wearable sensor using deep
  learning for gait analysis,'' \emph{IEEE Sensors Letters}, vol.~10, no.~4,
  pp. 1--4, 2026, art. no. 5502004.

\end{thebibliography}

\end{document}